\documentclass[article]{aa}
\usepackage{dcolumn}
\newcolumntype{d}{D{.}{.}{-1}}

\usepackage{natbib}
\bibpunct{(}{)}{;}{a}{}{,}
\usepackage{graphicx}
\usepackage{revsymb}	
\usepackage{amsmath}	
\usepackage[usenames]{color}

\newcommand{\msol} {$M_{\odot}$\,}

\usepackage{txfonts}
\usepackage[english]{babel}

\usepackage[switch]{lineno} 

\begin{document}

\title{Seismic analysis of HD\,43587Aa,\\ a solar-like oscillator in a multiple system}

\author{P. Boumier\inst{1}, O. Benomar\inst{2}, F. Baudin\inst{1}, G. Verner\inst{3}, T. Appourchaux\inst{1}, Y. Lebreton\inst{4,5}, P. Gaulme\inst{6}, W. Chaplin\inst{7}, R.A. Garc\'\i a\inst{8}, S.~Hekker\inst{9,10}, C. Regulo\inst{11,12}, D. Salabert\inst{13}, T. Stahn\inst{14}, Y. Elsworth\inst{7}, L. Gizon\inst{10,14}, M. Hall\inst{3}, S.~Mathur\inst{15,16}, E. Michel\inst{17}, T.~ Morel\inst{18}, B.~Mosser\inst{17},  E. Poretti\inst{19}, M. Rainer\inst{19}, I. Roxburgh\inst{3}, J.-D.~do~Nascimento~Jr.\inst{1,20}, R.~Samadi\inst{17}, M.~Auvergne\inst{17}, S.~Chaintreuil\inst{17}, A. Baglin\inst{17}, C. Catala\inst{17}}

\institute{
Institut d'Astrophysique Spatiale, CNRS, Universit\'e Paris Sud, (UMR 8617), 91405 Orsay Cedex, France
\and
Sydney Institute for Astronomy (SIfA), School of Physics, University of Sydney, New South Wales 2006, Australia
\and
Astronomy Unit, Queen Mary, University of London, Mile End Road, London E1 4NS, UK
\and
GEPI, Observatoire de Paris, CNRS (UMR 8111), 92195 Meudon, France 
\and
 Institut de Physique de Rennes, Universit\'e de Rennes 1, CNRS (UMR 6251), 35042 Rennes, France 
\and
New Mexico State University, Department of Astronomy, P.O. Box 30001, MSC 4500, Las Cruces, NM, USA
\and
School of Physics and Astronomy, University of Birmingham, Edgbaston, Birmingham B15 2TT, UK
\and
Laboratoire AIM, CEA/DSM-CNRS, Universit\'e Paris 7 Diderot, IRFU/Sap, Centre de Saclay, F-91191 Gif-sur-Yvette, France
\and
Astronomical Institute ``Anton Pannekoek'', University of Amsterdam, Science Park 904, 1098 XH, Amsterdam, the Netherlands
\and
Max-Planck-Institut f\"ur Sonnensystemforschung, Justus-von-Liebig-Weg 3, 37077 G\"ottingen, Germany
\and
Instituto de Astrof\'isica de Canarias, 38205, La Laguna, Tenerife, Spain
\and
Universidad de La Laguna, Dpto de Astrof\'isica, 38206, La Laguna, Tenerife, Spain 
\and
Universit\'e de Nice Sophia-Antipolis, CNRS, Observatoire de la C\^ote d'Azur, BP 4229, 06304 Nice Cedex 4, France
\and
 Institut f\"ur Astrophysik, Georg-August-Universit\"at G\"ottingen, 37077 G\"ottingen, Germany
\and
High Altitude Observatory, NCAR, P.O. Box 3000, Boulder, CO 80307, USA
\and
Space Science Institute, 4750 Walnut Street Suite 205, Boulder, CO 80301, USA
\and
LESIA, Observatoire de Paris, CNRS, Universit\'e Paris 6, Universit\'e Paris 7, (UMR 8109), F-92195 Meudon Cedex, France
\and
Institut d'Astrophysique et de G\'eophysique, Universit\'e de Li\`ege, All\'ee du 6 ao\^ut, B\^at. 5c, 4000 Li\`ege, Belgium
\and
 INAF-Osservatorio Astronomico di Brera, Via E. Bianchi 46, 23807 Merate, Italy 
\and
 Depart. de F\'isica Te\'orica e Experimental, Univ. Federal do Rio
   Grande do Norte, CEP: 59072-970 Natal, RN, Brazil 
}

   \offprints{P. Boumier}
   \mail{patrick.boumier@ias.u-psud.fr}
   \date{\today}

  \authorrunning{Boumier et al.}
  \titlerunning{Seismic analysis of HD\,43587Aa}

\abstract
{The object HD\,43587Aa is a G0V star observed during the 145-day LRa03 run of the COnvection, ROtation and planetary Transits space mission (CoRoT), for which complementary 
High Accuracy Radial velocity Planet Searcher (HARPS) spectra with S/N$>$300 were also obtained. Its visual magnitude is 5.71, and its effective temperature is close to 5950\,K. It has a known companion in a highly eccentric orbit and is also coupled with two more distant companions.}{We undertake a preliminary investigation of the internal structure of HD\,43587Aa.}{We carried out a seismic analysis of the star, using maximum likelihood estimators and Markov Chain Monte Carlo methods.}
{We established the first table of the eigenmode frequencies, widths, and heights for HD\,43587Aa. The star appears to have a mass and a radius slightly larger than the Sun, and is slightly older ($5.6$ Gyr). Two scenarios are suggested for the geometry of the star: either its inclination angle is very low, or the rotation velocity of the star is very low.}
{A more detailed study of the rotation and of the magnetic and chromospheric activity for this star is needed, and will be the subject of a further study. New high resolution spectrometric observations should be performed for at least several months in duration.}


\maketitle

\section{Introduction}
\label{intro}
The star HD\,43587Aa is a primary target of the asteroseismology program of the COnvection, ROtation and planetary Transits space mission (CoRoT)\footnote{The CoRoT space mission, launched on 2006 December 27, was developed and is operated by the Centre national d'études spatiales (CNES) with participation of the Science Programs of the European Space Agency (ESA), ESA's Research and Scientific Support Department (RSSD), Austria, Belgium, Brazil, Germany and Spain.}. CoRoT observed ten main sequence stars displaying solar-like oscillations \citep{Michel08}. This set of pulsators consists of: five F-stars, HD\,49933, HD\,181420, HD\,181906, HD\,175726, and HD\,170987 (see \citealt{Appourchaux08,Benomar2009b,Barban09,Garcia_2009,Mosser_2009,Mathur_2010}); the evolved G star HD\,49385  \citep{Deheuvels10}; the K and G stars HD\,46375 and HD\,52265, which host non-transiting planets (\citealt{Gaulme_2010,Ballot11}); the G star HD\,169392, which belongs to a weakly bound binary system \citep{Mathur_2013}; and the star that is the subject of the present study HD\, 43587Aa (G0V).\

Binary systems are fundamental astrophysical objects: when coupled with asteroseismology, they may provide two independent methods to obtain masses and radii, and therefore offer opportunities to test internal structure modelling and develop highly constrained stellar models. Indeed, it is occasionally possible to determine the masses and/or radii of stars constituting a binary or a multiple star system, provided that the orbital parameters are measurable. These can be obtained either by optically observing the components of visual binaries, or by combining photometric and radial velocity measurements in the case of eclipsing binaries. An alternative way of retrieving masses in photometry, for close-in binary systems, consists of modelling the photometric modulation resulting from the tidally distorted shapes of the stars, which are a function of stellar masses (e.g., \citealt{Mazeh_2008}). Finally, masses of triple system components, where at least two components mutually eclipse, can be deduced from eclipse timing variations (e.g., \citealt{Steffen_2013}). Recently, efforts were made to exploit the combination of binarity and asteroseismology, in particular with data from the NASA \textit{Kepler} mission. \citet{Hekker_2010} and \citet{Frand13} reported the identification of the eclipsing binary system KIC\,8410637, composed of an F star and a red giant that exhibit solar-like oscillations. \citet{Metcalfe_2012} performed the seismic analysis of the binary system 16 Cyg A and B, where eigenmodes of oscillations were identified for both solar-type stars. More recently, \citet{Gaulme_2013} reported the detection of 12 new eclipsing binary systems, one non-eclipsing binary with tidally induced oscillations, and ten more candidate triple systems, all of which include a pulsating red giant.\


\begin{figure}
\resizebox{\hsize}{!}{\includegraphics{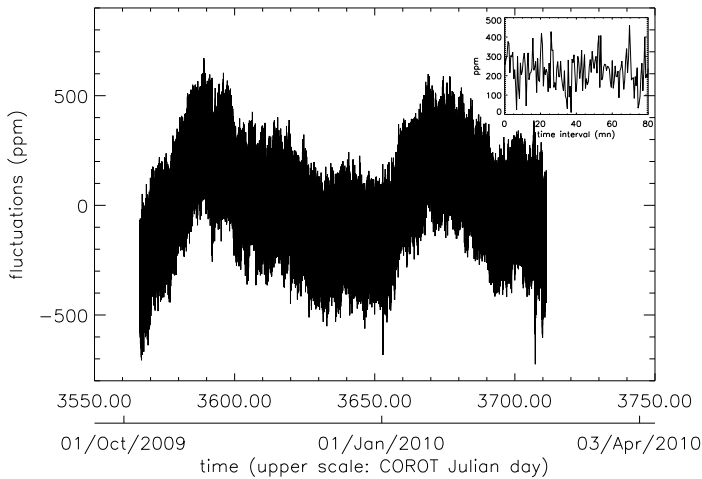}}
\caption{Light curve of HD\,43587 (LRa03 run of CoRoT), after a second-order polynomial fit was removed. A close-up of 80 minutes displays the temporal signature of the most powerful eigenmodes.}
\label{lightcurve}
\end{figure}


The subject of our work, HD\,43587Aa, a G0V star whose visual apparent magnitude is $m_V=5.71$, belongs to a quadruple system composed of two distant main sequence visual binaries (HD\,43587~A \& E) as imaged by \citet{Hart12} and by \citet{Pravdo06}, respectively. The denomination A and E is that of the Washington Double Star catalogue (WDS).
We were able to check that HD\,43587A and HD\,43587E had common proper motions and parallaxes measured with sufficient precision \citep{Lepi05}. Thanks to radial velocity measurements, \citet{Vogt02} were the first to determine estimates of the orbital properties of HD\,43587A, followed by \citet{Catala06} who used adaptive optics to resolve both stars in the system. Recently, \citet{Katoh13} found the most precise values to date using high-dispersion spectroscopy. The orbital period, eccentricity, and semi-major axis are estimated to be $P = 34.2 \pm 0.2$~ years, $e = 0.8045 \pm 0.0009$, and $a = 11.61 \pm 0.11$\,AU respectively. \citet{Katoh13} gave also estimates of $1.01\pm 0.02~M_\odot$ and $5926 \pm ~80$ K for the mass and the effective temperature of  HD\,43587Aa; they also derived a minimum value of $0.342\pm 0.003~M_\odot$ for the close companion HD\,43587Ab. The system HD\,43587A thus appears to be composed of a solar analog and an M star (effective temperature of $3820 \pm 100$ K following \citealt{Catala06}), and to orbit in a highly eccentric orbit.\\

A spectroscopic analysis was carried out by \citet{Bruntt04} who computed effective temperatures via different methods: 5923\,K (line depth ratio), 5850\,K (H$\alpha$ wings), and 5931\,K (Stromgren indices). Recently, \citet{Morel13} published a rather precise value, based on new data: $T_{\rm eff} = 5947\pm17$\,K  (more details are given in section 2.1).
\citet{Vogt02} measured $v\,\sin i$ to be 2.7\,km/s, while \citet{Bruntt04} suggested a value of 2.5\,km/s even though this is an extrapolation beyond the resolution limit of the spectrograph they used.
The star has a revised Hipparcos parallax of 52.0\,mas and a metallicity of $-0.11$ \citep[in the G-C catalogue and][]{Bruntt04}. \citet{Bruntt04} estimated the surface gravity to be 4.31 from Stromgren indices, and 4.29 from evolutionary tracks and parallax.
As far as the radius is concerned, a careful calculation was done by \citet{Thevenin06} using a calibrated Barnes-Evans relation to obtain a limb-darkening estimate and they found a value of $R = 1.280\pm0.032 R_{\odot}$.\

We will present in Section 2 the spectrometric ground-based support that was specifically performed to refine the temperature and abundance estimates, as well as the CoRoT photometric data that led to the identification of a high signal-to-noise ratio oscillation spectrum. Next, we describe how we measure the modes' properties (Sect. 3) and interpret them in terms of stellar bulk properties (Sect. 4). Finally, the modelling of the stars' internal structure with an evolutionary code allows us to estimate its parameters to an unprecedented precision (Sect. 5). For the sake of simplicity, HD\,43587Aa will be called HD\,43587 hereafter.


\begin{figure}
\resizebox{\hsize}{!}{\includegraphics{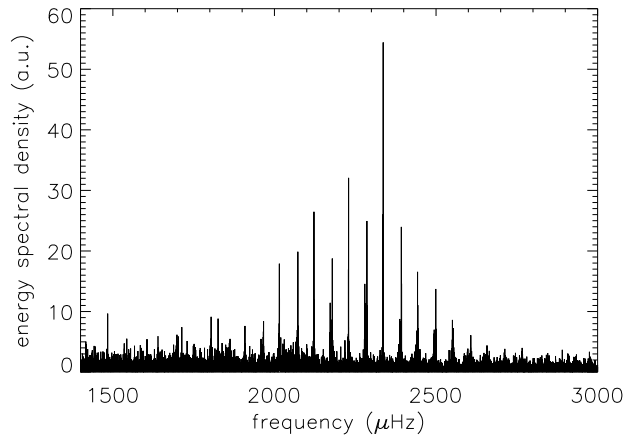}}	
\caption{Power spectrum of HD\,43587 in the p-mode range.}
\label{dse}
\end{figure}

\section{Observations and data}

\subsection{Spectroscopic data\protect\footnote{Based on spectroscopic observations made with the ESO 3.6m telescope at La Silla Observatory under the ESO Large Programme LP185.D-0056.}}


\begin{figure*}
\resizebox{\hsize}{!}{\includegraphics{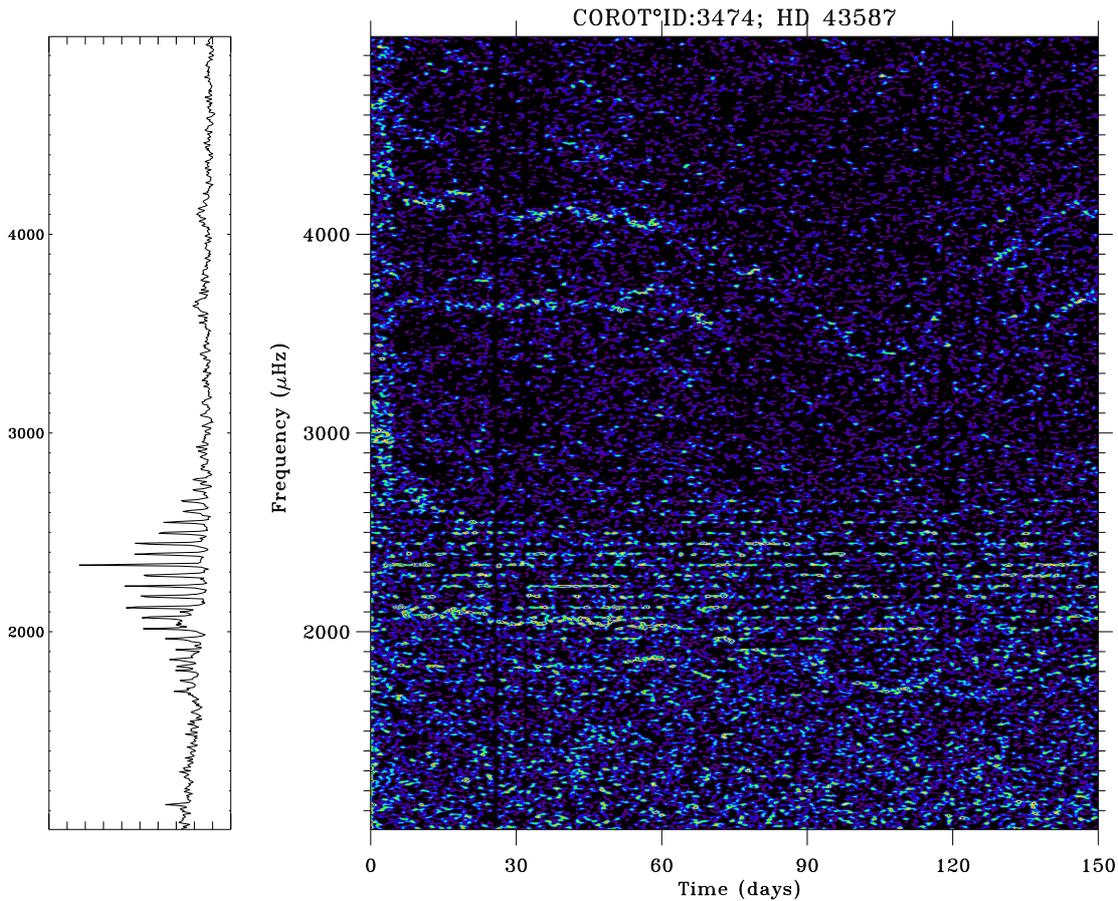}}
\caption{The time-frequency analysis of the time series of HD\,43587. Left: the collapsogram of the 2-D time-frequency image along the time axis (estimate of the mean spectrum).}
\label{TO}
\end{figure*}

New spectroscopic data were acquired to determine more accurate stellar parameters and abundance analysis. Three high signal-to-noise ratio spectra (S/N=410, 390, and 320) were obtained in December 2010 and January 2011 in the framework of the European Southern Observatory (ESO) LP185.D-0056. Solar spectra taken with the High Accuracy Radial velocity Planet Searcher (HARPS) were also used as reference for the subsequent local thermodynamic equilibrium abundance analysis, performed using the MOOG software and Kurucz atmospheric models with updated opacity distribution functions were used by \citet{Morel13} to analyse these data. MARCS models were also used for checking, leading to very minor differences. \citet{Morel13} obtained the following properties for HD\,43587:  $T_{\rm eff} = 5947\pm17$\,K, $\log g = 4.37\pm0.04$, and [Fe/H] $ = -0.02\pm0.02$. The uncertainties take statistical and systematic errors into account. The statistical errors are related to the uncertainties inherent in the properties of the numerous atomic lines used, extracted from four different lists: \citet{Reddy2003,Chen2000,Feltzing2001}; and \citet{Bensby2003}. As far as systematic errors are concerned, most sources were minimized by the fact that the analysis was strictly differential with respect to the Sun, which has well-known parameters (very close to HD\, 43587). This explains the low level obtained for the uncertainties of the stellar properties. Similar studies reported in the literature have claimed even lower (by a factor of 2) uncertainties, using this type of analysis for solar twins/analogues (\citealt{Mel12,Mon13}). \citet{Morel13} provided an accurate determination of elemental abundances, which confirms that the chemical composition of HD\, 43587 is close to solar, except for a slight magnesium excess and an enhancement by one order of magnitude of lithium (+2.05 dex for HD\, 43587 versus +0.92 dex for the Sun). This property will be discussed in Section 6, in light of the age of the star as derived from our asteroseismic analysis.\\

In addition, the rotational broadening and the macroturbulence have been estimated by fitting three isolated Fe\,I lines with [Fe/H], $v\sin i$, and $v_{\rm macro}$ as free parameters. The best fit values for the 3 Fe lines are $v\sin i \sim 0.4$\,km/s, and 3.3\,km/s\,$<v_{\rm macro}<$\,4.2\,km/s, typical of dwarfs with the temperature of HD\,43587 \citep[see Fig.\,3 of][]{Valenti2005}. It should be noted that performing the same analysis on a solar spectrum, the best-fit value for $v\sin i$ is not the true value of 1.8\,km/s but somewhat lower (although the true value is within the uncertainty). This shows that the $v\sin i$ value for HD\,43587 is very likely underestimated and that the determination of $v\sin i$ based on line-profile fitting is poorly constrained because of degeneracy problems \citep[see also][]{Bruntt2010}.

\subsection{CoRoT data}

We observed HD\,43587 with CoRoT during the third long run towards the galactic anti-center direction (LRa03), i.e., from October 2009 to March 2010. We performed the analysis on a 145-day  series, beginning once the geometrical mask used for the aperture photometry was optimised on the CCD detector. The first point of the series corresponds to the CoRoT Julian date 3566.2565. Fig.~\ref{lightcurve} shows the light curve extracted from the level 2 official product (so-called $helreg$ level 2 datasets, \citealt{Samadi07}), after removal of a second-order polynomial fit: the increasing trend in the raw light curve could come from the star itself but could also come from an overcorrection in the data reduction pipeline. The light curve corresponds to a 32-s sampled series, regularly spaced in the heliocentric frame, in which gaps corresponding to observations taken across the South Atlantic anomaly (SAA) were filled in with a linear interpolation. These gaps account for about 9\% of the original time series, which represents nearly the whole data loss (the duty cycle is close to 90\%). We checked that this interpolation did not produce any artifacts in the spectrum, the original orbital harmonics being instead significantly damped. Apart from low frequency orbital peaks, at 161.7 $\mu$Hz and its harmonics, the energy spectral density of the light curve exhibits very nice peaks, with the maximum of the power being around 2300 $\mu$Hz, as shown on Fig.~\ref{dse}.

Apart from the p-mode peaks, a hump of excess power is also visible just above 2000\,$\mu$Hz. The source of this power is attributed to instabilities of the image spot barycenter, as the spectrum of the position of the star on the CCD shows a similar hump. This is confirmed by a time-frequency analysis, which shows the variation of the main structures of the spectrum (Fig.~\ref{TO}): the horizontal ridges correspond to the p modes of the star, while the features fluctuating in frequencies are also observed in the horizontal and vertical positions of the spot barycenter. Moreover, such features are also seen in the data from other targets of the run. During the life of CoRoT, there are only a few runs for which several datasets suffer from such perturbations. We have not yet found what specifically produces such features in these runs, but we are currently conducting further work to understand this problem and correct it. Another star of the LRa03 run, HD\,43823, does not show any p-mode signal, but shows very clear trends similar to those of Fig.~\ref{TO}. The light curve of this star will be used in the following section, to discriminate between possible p modes and instrumental signals for HD\,43587. Note that Fig.~\ref{TO} presents the entire run, i.e., for the 150-day series. A vertical band of power excess shows that the signal was noisier during the first 4.5 days of the run, before the aperture mask was readjusted on the CCD. Finally, the power excess also appears near 4000 $\mu$Hz, but only the 2000 $\mu$Hz hump will make the modes parameters' estimation more difficult.

\begin{figure}
  \resizebox{\hsize}{!}{\includegraphics{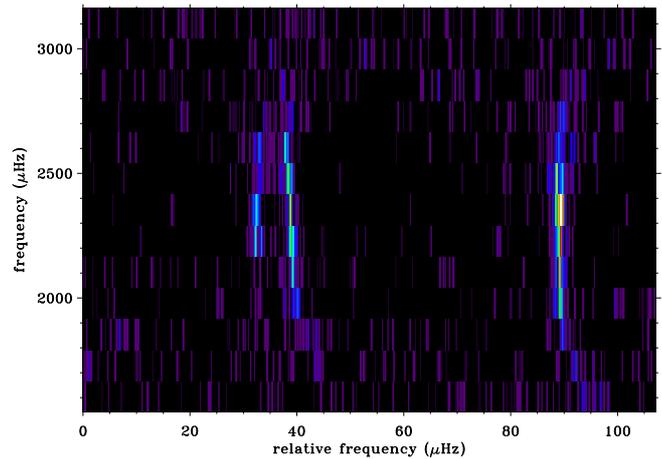}}
\caption{The echelle diagram of the power spectrum of HD\,43587 (folding of 107\,$\mu$Hz; 0.3 $\mu$Hz smoothing). The diagram is normalised by the local power, which increases the appearance of the noise at low and high frequency.}
\label{echel}
\end{figure}

As could be expected from Fig.~\ref{dse}, the echelle diagram shows distinct ridges, with a large separation of close to 107\,$\mu$Hz. These can be identified with modes of degree $\ell$\,=\,1, 2 and 0 (from left to right on Fig.~\ref{echel}). The other scenario, i.e., a ridge for $\ell=0$ and a split one for $\ell=1$ is rejected with a likelihood very close to 100\%, based on a likelihood ratio test, and supported by the fact that it would imply an internal rotation considerably larger than the projected surface rotation.

As far as the M star companion (HD43587Ab) is concerned, we have not detected any seismic signature in the power spectral density, which is not surprising given its faint magnitude (between 10 and 11, following, e.g., \citealt{ Catala06, Fuhrmann11, Hart12}).

\section{Determination of the frequency table}


\subsection{Determination of a reference frequency list}
Ten fitters produced p-mode parameter tables, using different methods: most of them (nine) used maximum likelihood estimators (MLE) based on methods developed by \citet{Appourchaux98}, while one used a bayesian approach based on Markov Chain Monte Carlo method \citep[MCMC; e.g.,] []{Benomar2009a}. Only four of them yielded results for some $\ell=3$ modes. As far as lower degree modes are concerned, in spite of different assumptions concerning, e.g., the stellar noise model and the relative amplitudes of the modes (either fixed or considered as free parameters), the consistency is excellent over a significant range of frequencies. In particular, within [1958; 2658] $\mu$Hz, the agreement is perfect (at 1-sigma level) between all the fitters for most of the modes. For six modes (out of 20) in this range, only one fitter is not in agreement with the others, the fitting process clearly having badly converged in these cases. For the specific $\ell=2$ case at 2065 $\mu$Hz, where the signal is contaminated by an instrumental signature, only five fitters yielded reliable estimates of the mode parameters; these five estimates are in agreement within 1-sigma.

Using only the frequencies obtained by all groups, we determined a reference table of frequencies with the following criteria:  
for each of these frequencies, we calculated the difference between the value obtained by each fitter and the mean value over all the fitters. This produced one series for each fitter. We chose the series with the minimum quadratic sum to be the reference, and we refer to the corresponding fitter as the reference fitter (RF). Frequencies and standard deviations obtained by the RF are displayed in Table~\ref{tablefreq2} in bold characters. The assumptions used by the RF are the following: one width per radial order was left as a free parameter (that of each radial mode with the $\ell$\,=\,2 and $\ell$\,=\,1 linewidth fixed to the nearest $\ell$\,=\,0 linewidth), and two heights per order were left free: the $\ell$\,=\,2 heights were fixed to 0.53 times the  $\ell$\,=\,0 height. This corresponds to the mean ratio found previously by the RF through a fit restricted to the highest peaks and allowing all the amplitudes to be determined independently. The rotational splitting and inclination angle were free parameters, and the background model consisted of two Harvey-like power-law components plus constant (white) noise. The fit was performed using classical maximum-likelihood optimisation. The errors were obtained from the inverse Hessian matrix of the fit. Note that modes labelled in Table~\ref{tablefreq2} with an exponent correspond to the modes correctly fitted by fewer than the ten fitters.

\subsection{Extension of the frequency table}
\begin{figure}
 \resizebox{\hsize}{!}{\includegraphics{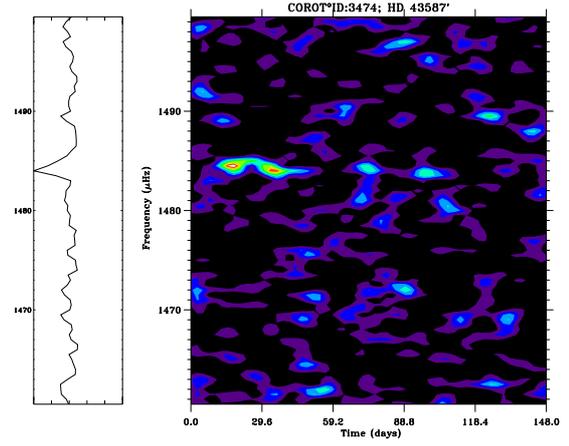}}	
\caption{Time-frequency diagram of HD\,43587 around 1484 $\mu$Hz.}
\label{TO1484}
\end{figure}

Outside of the frequency range above, we had a deeper look at each fitter's result, confronted with less homogeneity in the results due to the lower signal-to-noise ratio. Seven modes were identified consistently (within 1-sigma) by at least six fitters, including the RF. We also added the corresponding RF results in Table~\ref{tablefreq2}, labelled with the numbers of successful fitters. 
Four other modes were more questionable, and are placed within parentheses in the table. One of them, at around 2602 $\mu$Hz, was fitted by five of us, including the RF, with a high uncertainty but with results that are in agreement at the 1-sigma level. The $\ell$\,=\,1 mode around 2766 $\mu$Hz (five fitters) and the $\ell$\,=\,2 mode around 2708 $\mu$Hz (three fitters) were not fitted by the RF, and so we retained the results of the fitter whose set of frequencies leads to the minimum quadratic sum described in the previous subsection. Note that for both cases, the results of all fitters are in agreement at the 1-sigma level. The last case is special as only two fitters gave estimates for an $\ell$\,=\,1 mode, while the spectrum clearly shows peaks well above the level of the noise. To check whether these peaks are a signature of the contamination mentioned in Section 2, we performed a time-frequency analysis around this frequency, for both HD\,43587 and HD\,43823, the latter being also contaminated, as mentioned in Section 2.  Fig.~\ref{TO1484} clearly shows the signal, especially during the first 40-50 days. On the contrary, the same diagram for HD\,43823 does not show any feature, apart from a quasi-horizontal straight line at 1467 $\mu$Hz corresponding to a 24-hour alias of the $nine^{th}$ orbital harmonic. We, therefore, concluded that the signal comes from the star itself and not from the instrument, and retained a value rounded to 0.1 $\mu$Hz, compatible at 1-sigma with the two determinations. 

Note that some results were excluded due to a large inhomogeneity among the fitters. The case of an  $\ell$\,=\,1 candidate close to 1700 $\mu$Hz is interesting as it illustrates the difficulty to get reliable results through an automated process very well. Eight fitters gave results for this mode, which suggested strong support for this candidate. However, looking at the results in detail reveals that three different structures were the ``winners" of this research for the ($\ell$\,=\,1, n=14) p mode, i.e., the power spectral excess around this frequency being a large forest of peaks. A time-frequency analysis, performed again for both HD\,43587 and HD\,43823, shows that the patches of power responsible for this forest of spectral peaks are the same for both stars, which led us to reject the results for this ``mode."

\subsection{Are there any $\ell$\,=\, 3 modes?}

\begin{figure}
\resizebox{\hsize}{!}{\includegraphics{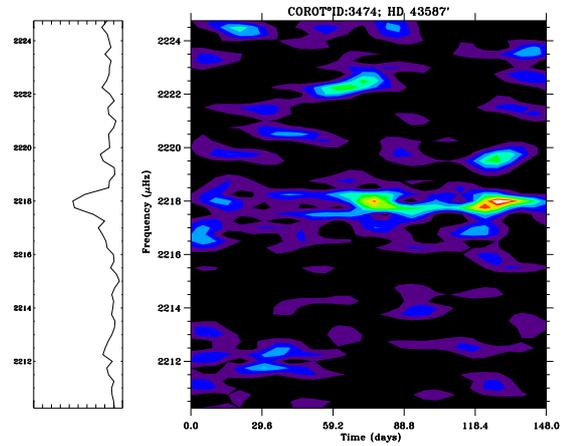}}
\caption{Time-frequency diagram of HD\,43587 around 2218 $\mu$Hz.}
\label{TO2218}
\end{figure}

The case of the $\ell$\,=\,3 modes was analysed specifically, as only four fitters, not including the reference fitter, have provided estimated frequencies for some of them. The highest peak that seems to have a significant signal-to-noise ratio is close to 2218 $\mu$Hz.  The fitters agree to better than 1-sigma, however, with formal uncertainties being a bit less consistent, ranging between 0.2 and 0.4 $\mu$Hz, which led us to compute the weighted mean. Fig.~\ref{TO2218} shows that the signal is significantly above the noise (signal not seen in  HD\,43823). At lower frequencies, the results are much less consistent, and even when the four values agree within 1-sigma, the uncertainties are too large, and the results are not reliable. At higher frequencies, there are results for three other fitted modes, which show an agreement well below the 1-sigma level, with reasonable uncertainties despite a low signal-to-noise ratio. Note that time-frequency analysis does not lead to results as clear as in the previous case. For consistency with lower $\ell$ cases, we put the last three estimates between parentheses as a warning about the low signal level. Note that analyses of the amplitudes of the modes (see next section) confirmed the presence of $\ell$\,=\,3 modes in the data.

\begin{figure}
\vspace{1.0cm}
\resizebox{\hsize}{!}{\includegraphics[angle=90]{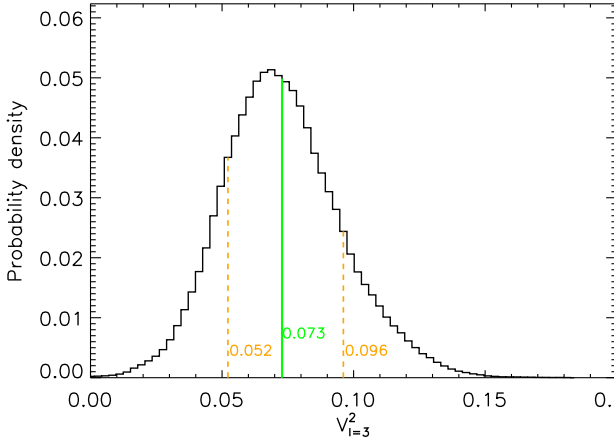}}
\caption{Probability density function of the relative height of the $\ell$\,=\,3 modes versus radial modes.}
\label{PDF_Height}
\end{figure}

\subsection{Width and amplitude}
The spectral height and width of the radial modes obtained by the reference fitter are given in Table~\ref{tablefreq2}. Note that the amplitude of the ($\ell$\,=\,0, n=18) was not considered reliable due to the 2 mHz hump of excess power (see section 2.2). 

In order to check the reliability of our estimates of the $\ell$\,=\,3 frequencies, the probabilities provided by a Bayesian approach are very helpful. This has been widely used in asteroseismology for several years now to fit stellar power spectra  (see, e.g., \citealt{Brewer2007, Benomar2009a, Gruberbauer2009, Benomar2009b, Handberg2011, Campante2011, Mathur2011, Benomar2012, Benomar2013}). Coupled with the MCMC sampling algorithm, a Bayesian approach allows us to extract accurately the full probability density function (PDF) of the parameters of the fit. Thus, contrary to maximization approaches (such as MLE), the uncertainties estimation is straightforward. Determining the 1-sigma confidence interval requires only the computation of the cumulative distribution function. Moreover, the samples from the MCMC process can be used to compute a PDF of any function of the parameters. As far as amplitudes are concerned, the PDF derived for the  $\ell$\,=\,3 amplitude ratio versus the radial modes shows that the probability of having a null $\ell$\,=\,3 amplitude is very close to zero. Fig.~\ref{PDF_Height} illustrates this result, the PDF of the ratio $(height_{\rm\ell=3}/height_{\rm\ell=0})$ having a median value close to 0.07. Note that the MCMC process also shows that the introduction of the $\ell$\,=\,3 modes leads to more stable results in the mode parameters determination.

\begin{figure}[!t]
\hspace{-1.0cm}
\includegraphics[angle=0,width=10 cm,height=10.0cm]{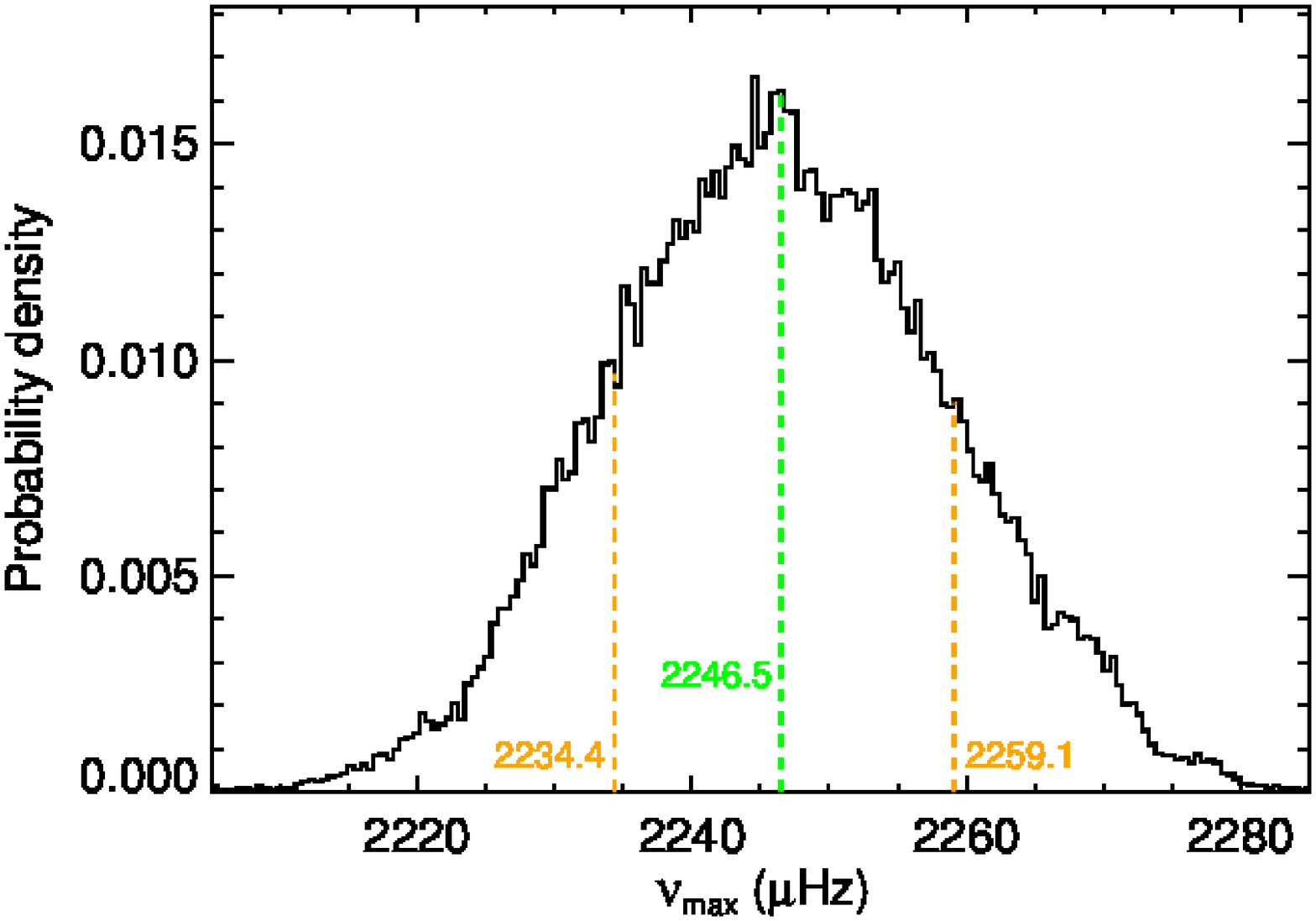}
\caption{Probability density function of the frequency of maximum oscillation power.}
\label{nu_max_Height}
\end{figure}

The frequency of maximum oscillation power ($\nu_{\rm max}$) was determined through an average over the frequencies weighted by the height of the modes, thus without any assumption as to the overall shape of the mode spectral envelope. The error propagation starting from MCMC results leads to the PDF shown in Fig.~\ref{nu_max_Height}, from which we derive $\nu_{\rm max} \sim 2247 \pm15\,  \mu $Hz. The maximum amplitude of the radial modes calculated with a cubic spline interpolation is $A_{\rm max} \sim 3.2\pm 0.6$ ppm.

 \begin{table*}
\caption{Table of frequencies, heights, and widths for HD\,43587. Bold values were used to determine the reference fitter. Frequency values in brackets were obtained by only a few fitters and have relatively high uncertainties.}              
\label{tablefreq2}      
\centering                                      
\begin{tabular}{c c c c c c c c c c c }          
\hline\hline                        
n&$\ell$&$v_{\rm \ell,n}$&height &width&$\ell$&$v_{\rm \ell,n}$&$\ell$&$v_{\rm \ell,n}$&$\ell$&$v_{\rm \ell,n}$\\    
&& [$\mu$Hz]&$ppm^{2}/\mu$Hz& $\mu$Hz&&[$\mu$Hz]&&[$\mu$Hz]&&[$\mu$Hz]\\
\hline
12&-&-&-&-&1&$(1484.4\pm 0.5)$\tablefootmark{2}&-&-&-&-\\
14&-&-&-&-&-&-&2&$1748.61\pm 0.38$\tablefootmark{6}&-&-\\
15&0&$1755.87\pm 0.22$&$1.8_{-0.3}^{+0.4}$&$1.3_{-0.6}^{+1.2}$&1&$1804.01\pm 0.19$\tablefootmark{7}&2&$1853.23\pm 0.28$\tablefootmark{6}&-&-\\
16&0&$1860.98\pm 0.20$&$3.1_{-0.5}^{+0.7}$&$1.2_{-0.5}^{+0.9}$&1&$1908.79\pm 0.19$\tablefootmark{8}&2&\boldmath{$1958.86\pm  0.28$}\tablefootmark{a}&-&-\\                          
17&0&\boldmath{$1965.95\pm 0.16$}&$3.0_{-0.6}^{+0.8}$&$1.3_{-0.5}^{+0.8}$&1&\boldmath{$2015.19\pm  0.15$}\tablefootmark{a}&2&\boldmath{$2065.8\pm  1.1 $}\tablefootmark{5}&-&-\\
18&0&\boldmath{$2072.41\pm  0.17$}&not reliable&$1.2_{-0.3}^{+0.4}$&1& \boldmath{$2122.20\pm  0.14$}&2&\boldmath{$2172.78\pm  0.33$}&3&$2217.9\pm 0.3)$\tablefootmark{4}\\
19&0&\boldmath{$2179.24\pm  0.13$}&$6.6_{-1.4}^{+1.7}$&$1.1_{-0.2}^{+0.3}$&1&\boldmath{$2229.26\pm  0.12$}&2&\boldmath{$2279.60\pm  0.14$}\tablefootmark{a}&3&$(2326.1\pm 0.5)$\tablefootmark{3}\\
20&0&\boldmath{$2286.08\pm  0.14$}&$5.5_{-1.1}^{+1.4}$&$1.3_{-0.4}^{+0.5}$&1&\boldmath{$2336.33\pm  0.11$}&2&\boldmath{$2387.37\pm  0.17$}\tablefootmark{a}&3&$(2433.2\pm 0.7)$\tablefootmark{4}\\
21&0&\boldmath{$2392.68\pm  0.14$}&$5.2_{-1.0}^{+1.3}$&$1.3_{-0.5}^{+0.8}$&1&\boldmath{$2443.05\pm  0.13$}&2&\boldmath{$2494.20\pm  0.21$}&3&$(2540.5\pm 0.9)$\tablefootmark{4}\\
22&0&\boldmath{$2499.24\pm  0.17$}&$3.4_{-0.7}^{+0.9}$&$1.4_{-0.5}^{+0.7}$&1&\boldmath{$2550.58\pm  0.21$}&2&$(2602.6\pm 0.6)$\tablefootmark{5}&-&-\\
23&0&\boldmath{$2607.01\pm  0.40$}&$0.8_{-0.2}^{+0.2}$&$2.6_{-1.1}^{+2.0}$&1&\boldmath{$2658.00\pm  0.42$}&2&$(2708.3\pm 1.4)$\tablefootmark{3}&-&-\\
24&0&$2713.1\pm 1.0$&$0.2_{-0.1}^{+0.1}$&$4.4_{-2.7}^{+7.0}$&1&$(2766.5\pm 0.8)$\tablefootmark{5}&-&-&-&-\\
\hline   \hline                        
\end{tabular}\\

\tablefoottext{a}{One fitter did not converge to the quoted value.}
\tablefootmark{2} \tablefootmark{3} \tablefootmark{4} \tablefootmark{5} \tablefootmark{6} \tablefootmark{7} {: indicates the number of fitters.}
\end{table*}

\section{Forward interpretation of mode properties}
\subsection{Frequency spacings}

From the table of eigenfrequencies we can derive the large separations, defined as:
\begin{equation}
\Delta \nu_{\rm \ell,n} =\nu_{\rm \ell,n} - \nu_{\rm \ell,n-1}.
\end{equation}

\begin{figure}
\resizebox{\hsize}{!}{\includegraphics{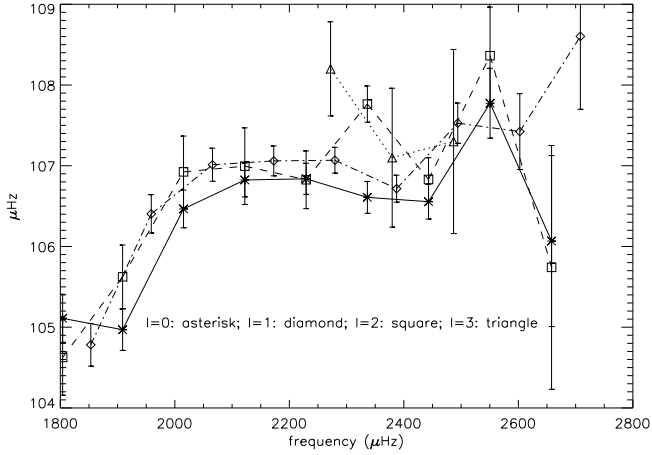}}
\caption{Large separations in frequency, for each degree.}
\label{lsep}
\end{figure}

Fig.~\ref{lsep} displays large separations obtained for each $\ell$\ separately. The mean value is close to 107\,$\mu$Hz, as could be already derived from Fig.~\ref{echel}, with a value of $\Delta \nu_{\rm max}=106.8\,\pm 0.2\,\mu$Hz around $\nu_{\rm max}$, for the radial modes. This quantity is linked to global properties of the star, such as the mass and the radius, as will be shown in the next subsection. Fig.~\ref{lsep} also shows very well that the radial modes ($\ell$=0) have a spacing significantly lower than for the higher $\ell$, as is the case for the Sun. This property can be better exploited in terms of structure diagnosis through the small frequency separations that potentially probe the stellar core itself. Small spacings are defined as (see, e.g., \citealt{Rox09}): 

\begin{equation}
\delta \nu_{02,n} =\nu_{\rm 0,n} - \nu_{\rm 2,n-1}
\end{equation}
\begin{equation}
\delta \nu_{\rm 01,n} =\nu_{\rm 0,n} - \frac{(\nu_{\rm 1,n-1}+\nu_{\rm 1,n})}{2}
\end{equation}
\begin{equation}
\delta \nu_{\rm 10,n} = \frac{(\nu_{\rm 0,n}+\nu_{\rm 0,n+1})}{2} - \nu_{\rm 1,n}.
\end{equation}

\begin{figure}
\resizebox{\hsize}{!}{\includegraphics{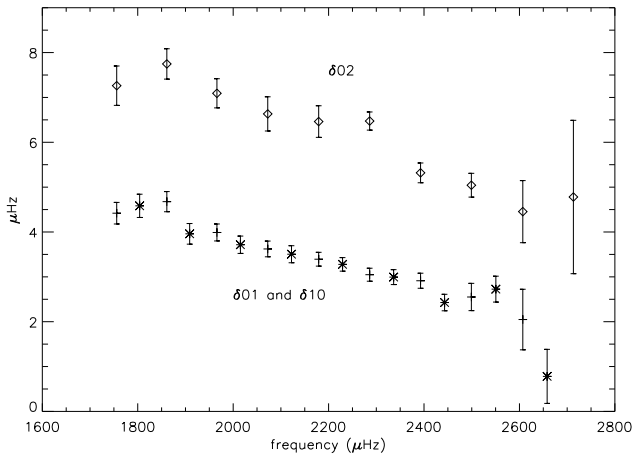}}
\caption{Small separations in frequency, between $\ell$\,=\,0 and $\ell$\,=\,2, and between $\ell$\,=\,0 and $\ell$\,=\,1.}
\label{ssep}
\end{figure}

Even if the large separation is somewhat smaller than the solar one, the small separations are close to those in the Sun and display a similar decrease with frequency (see Fig.~\ref{ssep}). A deeper analysis is needed to investigate any oscillation superimposed on the main trend.

\subsection{Estimates of global properties of the star}

Scaling relations are now commonly used to infer global characteristics of the star from its asteroseismic quantities (see, e.g., \citealt{Kjeld95} and \citealt{Belkacem11}). They link the mass and the radius of the star to $\nu_{\rm max}$, the frequency of the maximum oscillation power, and to $\Delta \nu_{\rm \nu_{\rm max}}$ the large frequency spacing close to $\nu_{\rm max}$, all quantities being scaled to solar values. Recently, \citet {Mosser13} proposed a revised version of the scaling relations, taking a bias coming from the difference between the observed and second-order asymptotic values of the large separation into account. The revised relations scale $\mathbf {\Delta \nu_{\rm as}}$, the asymptotic value of $\Delta \nu$, to calibrated solar reference values. The parameter $\Delta \nu_{\rm as}$ is the frequency difference between modes at high frequency, which can be estimated from $\nu_{\rm max}$ and $\Delta \nu_{\rm max}$, as shown by Eqs. (19) and (20) of \citet {Mosser13}:

\begin{equation}
\Delta \nu_{\rm as} = (1 + 0.57 \frac{ \Delta \nu_{\rm \nu_{\rm max}}}{\nu_{\rm max}})  \Delta \nu_{\rm \nu_{\rm max}}.
\end{equation}

The scaling relations, equivalent to Eqs. (28) and (29) of \citet {Mosser13}, can be expressed as follows:

\begin{equation}
\frac{M}{M_{\odot}} = \frac{ (\nu_{\rm max} / \nu_{\rm ref})^3 (T_{\rm eff} / T_{\rm eff \odot})^{3/2} } {( \Delta \nu_{\rm as} / \Delta \nu_{\rm ref})^4}
\end{equation} and

\begin{equation}
\frac{R}{R_{\odot}} = \frac{ (\nu_{\rm max} / \nu_{\rm ref}) (T_{\rm eff} / T_{\rm eff \odot})^{1/2} } {( \Delta \nu_{\rm as} / \Delta \nu_{\rm ref})^2}.
\end{equation}
\\
Using $\nu_{\rm max}= 2247\pm15\,  \mu$Hz, $\Delta \nu_{\rm \nu_{\rm max}}=106.8\,\pm 0.2\,\mu$Hz from our analysis, $T_{\rm eff} = 5947\pm17$\,K from \citet{Morel13}, and with $\nu_{\rm ref}= 3104\,\mu$Hz, and $\Delta \nu_{\rm ref}=138.8\,\mu$Hz from \citet {Mosser13}, we obtained a seismic estimation of the mass and of the radius of  HD\,43587:

\begin{equation}
M = 1.02 \pm 0.08  M_{\odot} 
\end{equation} and
\begin{equation}
R = 1.18 \pm 0.05  R_{\odot}.
\end{equation}

The uncertainties quoted above are not internal errors, but are based on the results of \citet{Mosser13}: their analysis shows that the performance of the calibrated scaling relations is about 4\% and 8\% for estimating, respectively, the stellar radius and the stellar mass, for masses less than 1.3 \msol. Taking this into account leads to error bars physically more realistic than those obtained in a straightforward approach by the propagation of the errors through the scaling relations, the latter being about a factor of two lower.

A third relation derived from the two above gives an estimate of the gravity, scaled to the Sun:

\begin{equation}
 \mathrm{log}\,g  =  \mathrm{log}\,g_{\odot} + \mathrm{log} \frac{\nu_{\rm max}} {\nu_{\rm max \odot}} + \frac{1}{2}  \mathrm{log} \frac{T_{\rm eff}}{T_{\rm eff \odot}}.
\end{equation}

 This yields the estimated value of log$\,g = 4.31\, $dex$ \pm 0.02$ (log$\,g_{\odot}= 4.44$).

\subsection{Determination of the rotation of the star}

\begin{figure}
\vspace{0.0cm}
\resizebox{\hsize}{!}{\includegraphics[angle=0,width=16.3cm,height=11.5cm,scale=1.2]{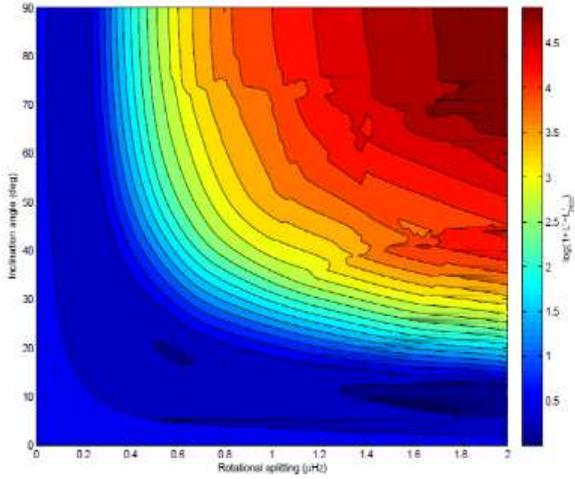}}
\\
\\
\caption{Likelihood map for rotational splitting and inclination. Lower (blue) values of L' correspond to better fits.}
\label{tetarotmap}
\end{figure}

\begin{figure}
\vspace{0.5cm}
 \resizebox{\hsize}{!}{\includegraphics[angle=90,width=14.3cm,height=10.0cm]{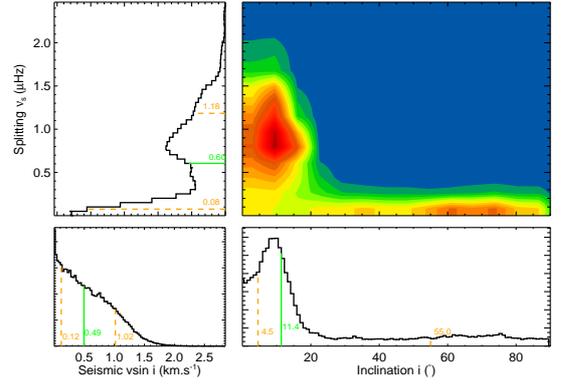}}
\caption{Correlation map of the rotational splitting with the inclination of the star. The projection on each axis gives the PDF of the rotational splitting (upper left) and of the inclination (lower right). The lower left panel represents the projected splitting.}
\label{PDF_projsplit}
\end{figure}

The internal rotation of the star could, in principle, be estimated by the splitting it produces on the spectral peaks of the modes. However, in the spectral density of HD\,43587, the peaks do not show a large amount of rotational splitting, and though the signal-to-noise ratio is high, none of the fitters could get reliable determinations of the splitting and of the inclination angle of the star. Fig.~\ref{tetarotmap} displays the likelihood map for rotational splitting and inclination obtained by the reference fitter. Unfortunately, there is no favored region in this 2-D map, so that many combinations of splitting and inclinations are possible. We, thus, tried to find the most probable solution from the PDF of each parameter determined by the MCMC algorithm. Fig.~\ref{PDF_projsplit} shows the bimodality of the PDF of the rotational splitting. The PDF of the inclination of the star does not help to discriminate this ambiguity: in spite of a clear peak at low values of the angle, the probability distribution has too large a spread in angle. The lower-left panel of Fig.~\ref{PDF_projsplit} displays the PDF of the seismic estimate of the projected splitting, $v\sin(i)$, derived from both splitting-inclination PDFs and from that of the radius estimated by the scaling law. This quasi-monotonically decreasing PDF confirms the difficulty we had reaching a conclusion and favours rather a small value of $v\sin(i)$, be  $\sim 0.5$\,km/s (median value) consistent with the spectroscopic reported in section 2.1 \citep{Morel13}. Two configurations are possible: either the star is seen at a very low inclination, and the rotational splitting is close to $1 \,\mu$Hz, or the splitting is too small to be resolved in the spectrum. For the same reason, spot modelling was also ineffective in deriving the mean rotation period \citep[]{Mosser09}. We will come back to the rotation and to the magnetic activity of the star in Section 6, in view of the modelling results of the star.

\section{Modelling the internal structure}

We performed modelling of the internal structure and evolution of the star with the public stellar evolution code Cesam2k \citep{Morel08}. We used the LOSC adiabatic pulsation code \citep{Scuflaire08} to compute the frequencies corresponding to a given internal structure model. A Levenberg-Marquardt minimization was applied to adjust the free parameters in the modelling to minimize the 
sum of the squared differences between observed and computed values \citep[see][]{Miglio05}:
\begin{equation}
\chi^2 = \sum_j \left( \frac{X^{\rm obs}_j - X^{\rm mod}_j}{\sigma_j} \right)^2
\label{eq:chi2}
\end{equation}
where $X_j$ are the $j$ observables ($\rm obs$ and $\rm mod$ specifying the observed or the modelled value) and $\sigma_j$ the associated uncertainty. 
In the present case, the models have been constrained by the effective temperature, surface metallicity, luminosity, and the individual oscillation frequencies. We considered the $35$ frequencies listed in Table\,\ref{tablefreq2}. We derived the luminosity $L$ from the $V$-magnitude given in the SIMBAD database, Hipparcos parallax \citep{vanLeeuwen07}, and the bolometric correction calculated according to \citet[][]{Vandenberg03}. As for the effective temperature $T_{\rm eff}$ and surface metallicity $\mathrm{[Fe/H]_s}$, we considered the spectroscopic data recently obtained by \citet{Morel13}. 
The adopted values of $T_{\rm eff}$, $L$, and $\mathrm{[Fe/H]_s}$ are listed in Table\,\ref{tab:mod_constr}.

In the modelling, the free parameters were the age of the
star, its mass, the initial helium abundance $Y_0$ in mass fraction, the initial ratio $(Z/X)_0$ of
metals to hydrogen, in mass fraction, and a free parameter $\alpha_\mathrm{CGM}$
entering the treatment of convection. We related the observed $\mathrm{[Fe/H]}$ to the present $(Z/X)$ ratio of HD~43587  through $\mathrm{[Fe/H]}=\log(Z/X)-\log(Z/X)_\odot$. As for the solar $(Z/X)_\odot$, we took the value $(Z/X)_\odot=0.0244$ from the canonical {\small GN93} solar mixture  \citep{Grevesse93}. In the calculations and in the opacity tables, all the individual abundances of metals are taken from Grevesse et Noels (1993).\\

More generally, in the modelling, convection was described following \citet{Canuto96} and microscopic diffusion, including gravitational settling, thermal, and concentration diffusion but no radiative levitation, was included following \citet{Michaud93}. We used the Nuclear Astrophysics Compilation of REaction rates (NACRE, \citealt{Angulo99}) except for the $^{14}$N$(p, \gamma)^{15}$O reaction where we adopted the revised Laboratory Underground Nuclear Astrophysics (LUNA, \citealt{Formicola04}). We also used the OPAL05 equation of state \citep{Rogers02}, the OPAL96 opacities \citep{Iglesias96} complemented at low temperatures by WICHITA tables \citep{Ferguson05} and the mixture of \citet{Grevesse93}. Near-surface effects on the frequencies were corrected following \citet{Kjeldsen08} and following the prescription of \citet{Brandao11}. 
The corrected frequencies $\nu_{n, l}^\mathrm{mod, corr}$ are calculated as:
\begin{equation}
\nu_{n, l}^\mathrm{mod, corr} = \nu_{n, l}^\mathrm{mod}+\frac{a}{r}\left(\frac{\nu_{n, l}^\mathrm{obs}}{\nu_\mathrm{max}}\right)^b
\end{equation}
\\
where $\nu_{n, l}^\mathrm{mod}$ and $\nu_{n, l}^\mathrm{obs}$ are, respectively, the model and observed frequencies, $b$ is an adjustable
coefficient, and $r$ approaches 1 as the model approaches the best solution. We treated $b$ as a variable
parameter of the modelling that we calibrated so as to minimize the differences between observed and computed individual
frequencies. The model presented here has $b=4$, $a=-5.8$, and $r=1.000$. In Eq. (11), the model frequencies $X_j^\mathrm{mod}$
are the corrected frequencies. Note that $\chi^2$  was calculated including seismic and non-seismic observations.\\

\begin{figure}
\resizebox{\hsize}{!}{\includegraphics[angle=0,width=14.3cm,height=10.0cm]{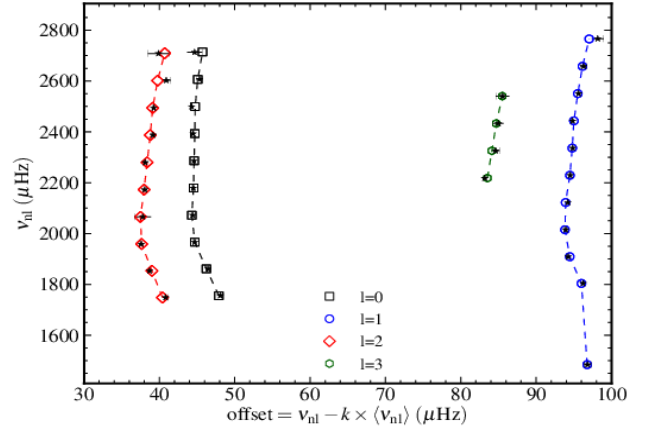}}

\caption{Echelle diagram of the best-fit model (with corrected frequencies). The observations are superimposed ($\star$ symbol with error bars).}
\label{echelcomp}
\end{figure}

Fig.~\ref{echelcomp} illustrates the high quality of the minimization procedure as far as the frequencies are concerned. It shows the echelle diagram of the best-fit model compared with the observations. The vast majority of modelled frequencies are compatible with the observed ones at the 1-sigma level. The mean values of the frequency separations, as defined by Eqs. (1) to (4), are listed in Table\,\ref{tab:sepcomp} for both observational and modelled cases. The precise inferences of the frequency dependence of these separations in terms of stellar internal structure are under study and will be the subject of a future paper (Lebreton and Goupil, in preparation). The main characteristics of HD\,43587 yielded by the modelling are listed in Table\,\ref{tab:mod_res}. Modelling of HD\,43587 converges to a star slightly more massive, hotter, and thus more luminous than the Sun: $M_\mathrm{model} = 1.04 M_{\odot};\, R_\mathrm{model} = 1.19R_{\odot};\,L_\mathrm{model}=1.58\,L_{\odot}$ and an age of $5.6$ Gyr. The base of the external convection zone is at $72.3$ percent of the stellar radius and has a temperature of $1.86\times 10^6$ K. It is worth pointing out that the mass and radius of the star built by the modelling are very close to the values yielded by the scaling ($M = 1.02 \pm 0.08  M_{\odot} , R = 1.18 \pm 0.05  R_{\odot}$). Note that one should consider the error bars on age and mass as internal errors, resulting from the Levenberg-Marquardt minimization for a given physical description of the stellar interior. The actual error bars, considering the uncertainties on the input physics of the models are expected to be higher, probably at the level of $10-15$ percent for the age, $2-4$ percent for the mass, and $1-2$ percent for the radius \citep{Lebreton12}. Another minimization performed using the Asteroseismic modelling Portal (AMP, \citealt{Metcalfe12}) led to an age of $5.66\,$Gyr, while the other parameters (mass, radius, temperature, luminosity) were found to be very close to those coming from the present minimization. \citet{Morel13} reported another estimate of the age, $4.94 \pm 0.53\, $Gyr, obtained with the PARAM isochrone inversion tool \citep{Girardi00, dasilva06}. It is worth pointing out that this age is based solely on grids of stellar evolutionary tracks without considering any asteroseismic constraints.

As far as gravity is concerned, the modelling yields $\log\,g = 4.31\, $dex, again in agreement with the value provided by scaling relations ($\log\,g=4.31\pm 0.02 $ dex). The value inferred from spectroscopic observations is slightly higher:  $\log\,g_\mathrm{spect} = 4.37\,  \pm 0.04$ dex \citep{Morel13}. Such a difference in the $\log\,g$ determinations was already observed by \citet{Bruntt12} on solar-type Kepler targets covering a wide range of effective temperatures, and by \citet{Mes13} on giant stars. Even though the asteroseismic determination is found to be model independent and more reliable \citep{{gai11},{Bruntt12},{Hek13}}, we recall that in the case of solar twins/analogues, the spectroscopic determination can be, in principle, obtained with a low uncertainty thanks to a strictly differential analysis with respect to the Sun. In any case, the systematic difference between the two log g determinations still needs to be understood.

\begin{table}
\center
\caption{Observational constraints considered in the modelling  of HD\, 43587. The error bars on  $T_{\rm eff}$ and $\mathrm{[Fe/H]_s}$ correspond to the $3 \sigma$ error bars in \citet{Morel13}.}
\label{tab:mod_constr}
\begin{tabular}{c|c|c}
\hline
\hline
 $T_{\rm eff}$ & $L$ & $\mathrm{[Fe/H]_s}$ \\
 (K) & $(L_{\odot})$ &  (dex) \\
\hline
$5947\pm51$ & $1.615\pm0.030$ & $-0.02\pm 0.06$ \\
\hline
\end{tabular}
\end{table}

\begin{table}
\center
\caption{Mean values over frequency for the frequency separations defined by Eqs. (1) to (4).}
\label{tab:sepcomp}
\begin{tabular}{c|c|c|c|c}
\hline
\hline
 &$<\Delta \nu>$ &$< \delta \nu_{02}>$ &$<\delta \nu_{\rm 01}>$&$<\delta \nu_{\rm 10}>$ \\
 &$(\mu \mathrm{Hz})$&$(\mu \mathrm{Hz})$&$(\mu \mathrm{Hz})$&$(\mu \mathrm{Hz})$ \\
\hline
Observations&$106.74\pm 0.11$&$6.12\pm 0.23$&$3.41\pm 0.07$ & $3.2 \pm 0.10 $ \\
\hline
Model&106.77 &6.33 &3.51 &3.39\\
\end{tabular}
\end{table}

\begin{table*}
\center
\caption{Characteristics of the best model. The age, mass, initial helium and metallicity, and convection parameter of the optimal model result from the Levenberg-Marquardt minimization. The corresponding values of the model radius, effective temperature, luminosity, $\mathrm{[Fe/H]_s}$ and the current convection zone helium abundance are also listed. }
\label{tab:mod_res}
\begin{tabular}{c|c|c|c|c|c|c|c|c|c}
\hline
\hline
 Age & Mass & $Y_0$ & $ (Z/X)_0$ & $\alpha_{\rm CGM}$  & Radius &$T_{\rm eff}$ & $L$ & $\mathrm{[Fe/H]_s}$ &$Y_{\rm cz}$\\
 (Gyr) & $(M_{\odot})$ &  &  & & $(R_{\odot})$ & (K) & $(L_{\odot})$ &  (dex) & \\
\hline
$5.60\pm0.16$ & $1.04\pm0.01$ & $0.288\pm0.006$ & $0.030\pm0.001$ & $0.68\pm0.02$ & $1.19$&$5951$&$1.583$& $0.01$ &0.243\\
\hline
\end{tabular}
\end{table*}

\section{Discussion}
HD\,43587Aa could appear as a big sister of the Sun, if not a solar twin, but looking deeper in detail shows that it is not so simple. We showed that the star is slightly older than the Sun, which adds to the difficulty in understanding why the lithium abundance is one order of magnitude higher than solar. Such an enrichment is unexpected at this age for this type of star (see, e.g., Fig. 7 of \citealt {Melendez10}), even when taking different initial conditions such as the rotation of the star into account (see Fig. 8 of \citealt {Melendez10} and Fig. 4.6 of \citealt {Marques13}). Only an extremely low initial rotation rate would explain the high lithium abundance, if only rotation were to be considered. As far as the rotation is concerned, our asteroseismic analysis does not give any value for HD\,43587Aa, but it shows that if the star is not pole on, then its rotation rate is probably significantly below that expected at this age. In any case, we should make it clear that a very small rotation of the star today would not necessarily explain the enhanced lithium abundance (thanks to a very small initial rotation), as after about 1 Gyr, the rotation does not depend on the initial conditions \citep {Marques13}. Unfortunately,  the analysis of the light curve for HD\,43587A does not yield the surface rotation period with certainty. The chromospheric behaviour of the Ca II H and K lines (S index), measured for the period of one month with the NARVAL spectropolarimeter gives an independent  estimate of the lower limit for HD\,43587Aa rotation, $P_{\mathrm {rot}}$ $>$ 15d. One should reobserve for several months to be able to determine the rotation satisfactorily.
Another interesting property of HD\,43587Aa is its very low chromospheric activity level (see, e.g., \citealt{hall09}). \citet{schroder12} even suggested that the star could be in a Maunder minimum period. \citet{lubin10} tried to connect the lithium abundance with the evolutionary state, the low activity, and a Maunder minimum phase, but the links between all of these factors are not straightforward and not fully understood.

\section{Conclusions}
The analysis of the light curve obtained by CoRoT of HD\,43587A exhibits clear acoustic modes, which allowed us to establish a reference frequency table. Time-frequency analysis and MCMC methods were used to help discriminate between noise and acoustic signal when the signal-to-noise ratio was low. The detection of several  $\ell$\,=\,3 modes  proved to be statistically significant. The frequency splitting induced by the internal rotation of the star could not be extracted satisfactorily, its probability density function being bimodal. Two scenarios could be realistic for the geometry of the star: either its inclination angle is very low, which disminishes the signature of rotation in the spectrum, or the rotation of the star is very low, with the same consequence on the spectrum. No conclusion could be reached concerning the surface rotation itself, with no magnetic structures due to the extremely low level of activity.

The HD\,43587Aa seismic model corresponds to a star slightly more massive, hotter, and thus more luminous than the Sun: $M_\mathrm{model} = 1.04 M_{\odot};\, R_\mathrm{model} = 1.19R_{\odot};\,L_\mathrm{model}=1.58\,L_{\odot}$ and an age of $5.6$ Gyr. The rotation profile remains particularly challenging: a solar analogue with a very slow rotation would result from a different history from that of the Sun. The physics that could be responsible for this (protostellar cloud, early age of the star, composition, magnetic field) remains uncertain, as well as the impact of the binarity of the HD\,43587A system on the evolution and the properties of HD\,43587Aa. New high resolution spectrometric observations should be realised with the sampling and the duration required to be able to decrease the upper limit of the rotation rate. A more precise analysis of the magnetic and chromospheric activity for this star is needed, and will be the subject of a further study.

\begin{acknowledgements}
PB wishes to thank J\'er\^ome Ballot, Kevin Belkacem, and Orlagh Creevey for fruitful discussions about this analysis. SH acknowledges financial support from the Netherlands organisation for Scientific Research and ERC starting grant $\#338251$ (Stellar Ages). TM acknowledges financial support from Belspo for contract PRODEX-GAIA/ /DPAC. EP acknowledges financial support from the 
 Italian PRIN-INAF 2010 {\it Asteroseismology: looking inside the stars 
with space- and ground-based observations}. MR acknowledges
   financial support from the FP7 project {\it SPACEINN: Exploitation of Space Data for
   Innovative Helio- and Asteroseismology}. IWR thanks the Leverhulme Foundation for support under grant EM-2012--35/4. RAG thanks the CNES grant at CEA. TS and LG acknowledge support from DFG Collaborative Research Center 963 “Astrophysical Flow Instabilities and Turbulence” (Project A18). This research has made use of the SIMBAD database,
operated at CDS, Strasbourg, France. We are also grateful to all of the people who made CoRoT possible and to those who are still working to ensure the best possible results. 

\end{acknowledgements}

\bibliographystyle{aa} 
\bibliography{HD43587_bib} 

\end{document}